\documentclass[preprints,article,communication,accept,moreauthors,pdftex,10pt,a4paper]{Definitions/mdpi} 

\firstpage{1} 
\pubvolume{5}
\issuenum{1}
\articlenumber{0}
\pubyear{2019}
\copyrightyear{2019}
\history{Received: 16 April 2019; Accepted: 22 May 2019; Published: date}

\Title{Jet Structure Studies in Small Systems$\dagger$}

\firstnote{\hangafter=1 \hangindent=1.05em \hspace{-0.82em}This paper is based on the talk at the 18th Zimányi School, Budapest, Hungary, 3–7 December 2018.}  

\Author{Zoltán Varga $^{1,2,}$*\orcidA{}, Róbert Vértesi $^{1}$\orcidB{} and Gergely Gábor Barnaföldi $^{1}$\orcidC{}}
\AuthorNames{Zoltán Varga, Róbert Vértesi and Gergely Gábor Barnaföldi}

\address{%
$^{1}$ \quad Wigner Research Centre for Physics, P.O. Box 49, H-1525 Budapest, Hungary; \mbox{vertesi.robert@wigner.mta.hu (R.V.)}; barnafoldi.gergely@wigner.mta.hu (G.G.B.) \\
$^{2}$ \quad Department of Theoretical Physics, Institute of Physics, Faculty of Science, Eötvös Loránd University, Pázmány Péter sétány 1/A, H–1117 Budapest, Hungary}

\corres{Correspondence: varga.zoltan@wigner.mta.hu}

\abstract{A study investigating a possible jet shape dependence on the charged event multiplicity was performed on collision samples generated by Monte--Carlo (MC) event generators {\sc Pythia} and {\sc Hijing++}. We calculated the integral jet shape and found a significant modification caused by multiple-parton interactions. By interchanging and enabling different model ingredients in the simulations and analyzing the results in several $p_\mathrm{T}$ bins and event multiplicity classes, we found a characteristic jet size measure that was independent of the chosen tunes, settings, and jet reconstruction algorithms.}



\begin{document}

\section{Introduction}

The discovery of collective-like behavior in high-multiplicity proton-proton (pp) and proton-nucleus (pA) collisions was one of the major surprises in early LHC
results~\cite{laa, gv}. The~collective-like behavior previously found in large systems, manifested in long-range correlations and a sizeable azimuthal anisotropy, have traditionally been considered as a signature proving the presence of the quark-gluon plasma (QGP).  The~creation of the QGP in high-energy nucleus-nucleus (AA) collisions can also be investigated by studying the structure of jets and their modification that leads to the well-known jet quenching phenomenon~\cite{Adcox:2001jp, Gyulassy:2000}. Whether QGP is created in small systems like pp collisions is still an open question~\cite{Dusling:2015gta}. However, the~presence of the QGP is not necessary to explaining collectivity: Relatively soft vacuum-QCD
effects such as multiple-parton interactions (MPI) and color reconnection (CR) can also produce a similar behavior~\cite{kpz, Ortiz:CR}. These interactions, at~least in principle, can modify the jet shapes in even small systems. Although~experimental confirmation is not yet available, a~recent phenomenology study also suggests the modification of hard processes by soft vacuum-QCD effects in a high-multiplicity environment~\cite{Mishra:2019png}.

MPI is also expected to depend on flavor~\cite{4jet}. Fragmentation of heavy-flavor jets is expected to differ from light-flavor jets because of color charge and mass effects. The~internal structures of heavy-flavor jets  may therefore provide a deeper insight into the flavor-dependent development of jets and their connection to the underlying event (UE). This paper continues our previous studies~\cite{VVB, VVB2} aimed at the evolution of jet structure patterns and their dependence on simulation~components.

\newpage

\section{Analysis}
We used {\sc Pythia} 8.226 and {\sc Hijing++} Monte--Carlo (MC) generators to simulate pp collision events at $\sqrt{s} = 7$ TeV~\cite{zea,jiz}. Three different {\sc Pythia} tunes were investigated, the~Monash 2013 tune with the NNPDF2.3LO PDF set~\cite{pea1,hta}, the~Monash* tune with NNPDF2.3LO~\cite{pea2}, and~tune~4C with the CTEQ6L1 PDF set~\cite{vw,ms}. Collisions with and without multiple-parton interactions and color reconnection were simulated and compared to each other~\cite{VVB,VVB2}.
We only considered particles above the transverse momentum threshold $p_\mathrm{T}^\mathrm{track}>0.15$ GeV/c. We carried out a full jet reconstruction using the $\mathrm{anti}$-$k_\mathrm{T}$, $k_\mathrm{T}$, and~Cambridge--Aachen jet reconstruction algorithms, which are part of the {\sc Fastjet} software package~\cite{ma}. These choices are typical in jet shape analyses~\cite{Chatrchyan:2012mec}.
The multiplicity-integrated jet shape studies of CMS
were used as a benchmark for our current multiplicity-differential studies~\mbox{\cite{VVB,Chatrchyan:2012mec}}. Therefore we chose the jet resolution parameter as $R=0.7$ and applied a fiducial cut so that the jets were contained in the CMS acceptance $|\eta| < 1$. We did not apply underlying event subtraction in the jet cones. We investigated the jets within the \mbox{15 GeV/c < $p_\mathrm{T}^\mathrm{jet}$ < 400 GeV/c} transverse momentum range, where multiplicity-differential studies on real data are feasible in the near~future.

We chose the following two jet shape measures to study the multiplicity-dependent behavior of the jet structure: The $\Psi$ integral jet shape (or momentum fraction) and the $\rho$ differential jet shape (or momentum density fraction)~\cite{jetshape1, jetshape2}. The~former one gives the average fraction of the jet transverse momentum contained inside a sub-cone of radius $r$ around the jet axis, the~latter one is the momentum profile of the jet, i.e.,~the average transverse momentum of the particles contained inside an annulus with a $\delta r$ width and boundaries $r_a$ and $r_b$. The~exact formulae are given by:
\begin{eqnarray}
\Psi(r) = \frac{1}{p_{\mathrm{T}}^{\mathrm{jet}}}\sum\limits_{\substack{r_\mathrm{i} < r}} p_{\mathrm{T}}^\mathrm{i}
& \mathrm{and} &
\rho(r) = \frac{1}{\delta r} \frac{1}{p_{\mathrm{T}}^{\mathrm{jet}}}\sum\limits_{\substack{r_a < r_\mathrm{i} < r_b}} p_{\mathrm{T}}^\mathrm{i} 
\end{eqnarray}
respectively, where $p_{\mathrm{T}}^\mathrm{i}$ is the transverse momentum of the selected particle and $p_{\mathrm{T}}^{\mathrm{jet}}$ is the transverse momentum of the jet. The~distance $r_\mathrm{i}$ of the given particle from the jet axis is calculated as \mbox{$r_\mathrm{i} = \sqrt{(\phi_\mathrm{i} - \phi_{\mathrm{jet}})^2 + (\eta_\mathrm{i} - \eta_{\mathrm{jet}})^2}$}, where $\phi$ is the azimuthal angle and $\eta$ is the pseudorapidity. For~a better understanding, we noted that the aforementioned observables are connected with the equations:
\vspace{-3mm}
\begin{eqnarray}
\Psi(r) = \int_0^r \rho(r') dr',
& \mathrm{and} &
\Psi(R) = \int_0^R \rho(r') dr' = 1,
\end{eqnarray}
where $R$ is the jet resolution parameter. The~differential and integral jet shapes were calculated for the above mentioned $p_\mathrm{T}^\mathrm{jet}$ range. The~three tunes reproduced the CMS results within statistical~errors~\cite{VVB,VVB2}.
\section{Results}

\begin{figure}[H]
\centering
\includegraphics[width=0.48\linewidth]{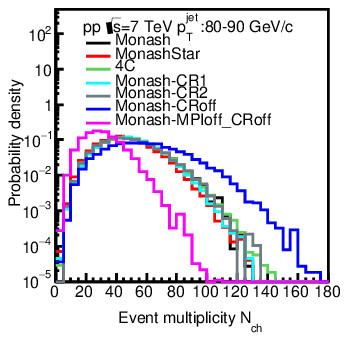}
\includegraphics[width=0.48\linewidth]{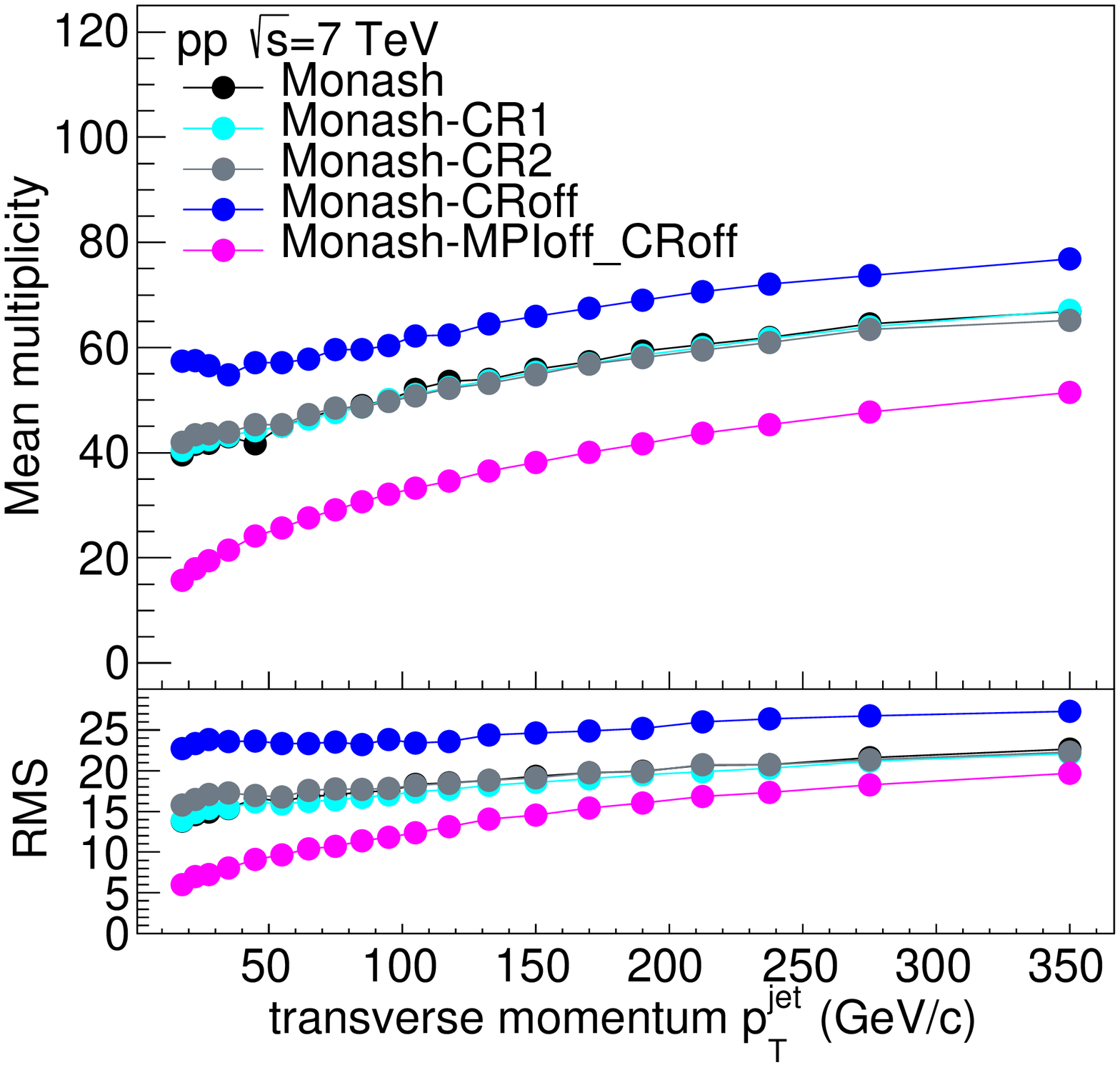}
\caption{\label{fig:mult}({\bf{Left}}) Multiplicity distributions for different tunes and settings in events where the $p_\mathrm{T}^\mathrm{jet}$ of the reconstructed jets falls in a given $p_\mathrm{T}$ bin; ({\bf{Right}}) The mean values of the distributions are shown as a function of the $p_\mathrm{T}^{\mathrm{jet}}$. The~RMS is the relative mean squared value of the~distribution.}
\end{figure}

The multiplicity distributions (multiplicity is defined as the number of charged final state particles in a given collision event) were very similar for the three tunes used in our simulations, as~shown in the left panel of Figure~\ref{fig:mult}. However, a~significant change is observed when we do not consider the effects of the color reconnection and/or the multiple-parton interactions. Without~the CR, the~multiplicity distribution becomes wider. If~we also switch off the MPI, the~multiplicity distribution becomes much narrower compared to the setting where both of them are applied. The~width of the multiplicity distribution, however, is not sensitive to the choice of the CR model. In~the right panel of Figure~\ref{fig:mult}, we plot the mean values of the multiplicity distributions in events where we reconstructed a jet of a given transverse momentum. We did not exclude non-leading jets in our analysis, but~we also investigate the effects of selecting only leading jets later in this paper. As~expected, the~average event multiplicity grew with the transverse momentum of the selected~jet.
\unskip

\begin{figure}[H]
\centering
\includegraphics[width=0.48\linewidth]{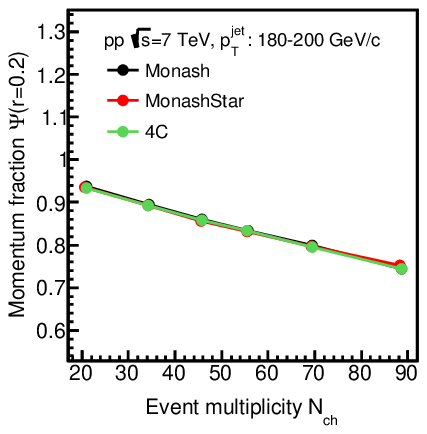}
\includegraphics[width=0.48\linewidth]{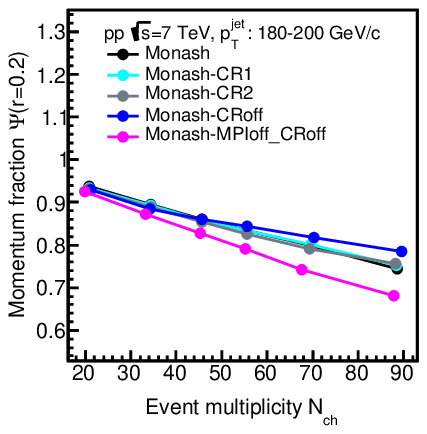}
\caption{\label{fig:psi} ({\bf{Left}}) Integrated jet shape for different tunes; ({\bf{Right}}) Integrated jet shape for different~settings.}
\end{figure}

The integral jet shape with $r=0.2$ is shown in Figure~\ref{fig:psi} to compare the effects of the different tunes and settings. As~expected~\cite{ATLAS:tune}, we see similar trends for the multiplicity distributions of the tunes. However, there is a substantial difference between the different MPI and CR settings. The~most significant difference in the jet shapes was caused by turning off the MPI, which further supports the current view that the MPI contributions need to be included to correctly describe the jet shapes. We note that we did not subtract the underlying event from the jet. Further investigation is necessary to understand whether the interplay between the UE and the observed hard process is significantly modified by the~MPI.

\begin{figure}[H]
\centering
\includegraphics[width=0.48\linewidth]{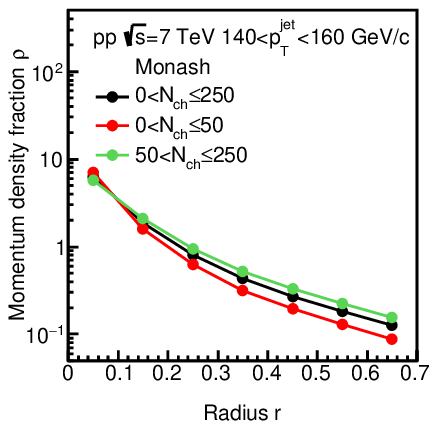}
\includegraphics[width=0.48\linewidth]{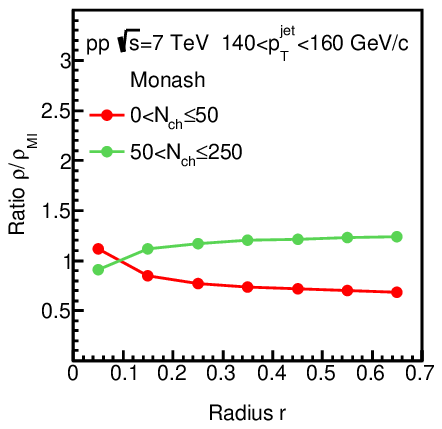}
\caption{\label{fig:rho} ({\bf{Left}}) Differential jet shape for low- and high-multiplicity bins; ({\bf{Right}}) The differential jet shape $\rho$ divided by the $\rho_\mathrm{MI}$
to emphasize the difference.}
\end{figure}

In order to gain a more comprehensive picture, we use the current
differential jet shape to investigate the multiplicity dependence of the jet shapes. On~the left side of Figure~\ref{fig:rho}, we categorize the events into a high and low multiplicity class. The~differential jet shapes of both categories are compared to the multiplicity-integrated momentum density ($\rho_\mathrm{MI}$), 
computed without any selection in multiplicity. A~multiplicity dependence is observed. The~jets in the higher multiplicity bin appear to be wider while the jets in the lower multiplicity bin appear to be narrower. This is a trivial multiplicity dependence since the event multiplicity strongly correlates with the jet multiplicity. In~our previous work, we canceled this trivial multiplicity dependence by applying a double ratio across the different~tunes~\cite{VVB}.

In the right panel of Figure~\ref{fig:rho}, we plotted the $\rho(r)/\rho_\mathrm{MI}$ ratio for the low- and high-multiplicty bins divided by the $\rho_\mathrm{MI}$ curve. In~this way, the~difference compared to the $\rho_\mathrm{MI}$ (black) curve is more visible. We also observed that the $\rho(r)/\rho_\mathrm{MI}$ curves obtained from the low- and high-multiplicty classes intersect each other at a particular radius inside the jet~cone.

\begin{figure}[H]
\centering
\includegraphics[width=0.48\linewidth]{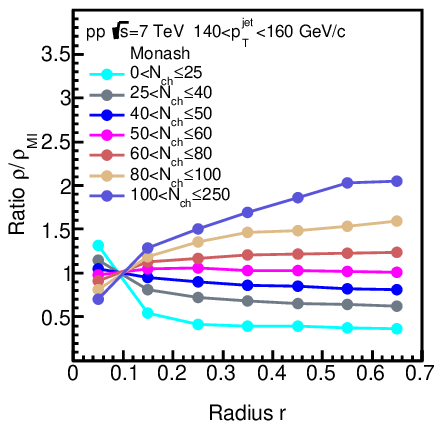}
\includegraphics[width=0.48\linewidth]{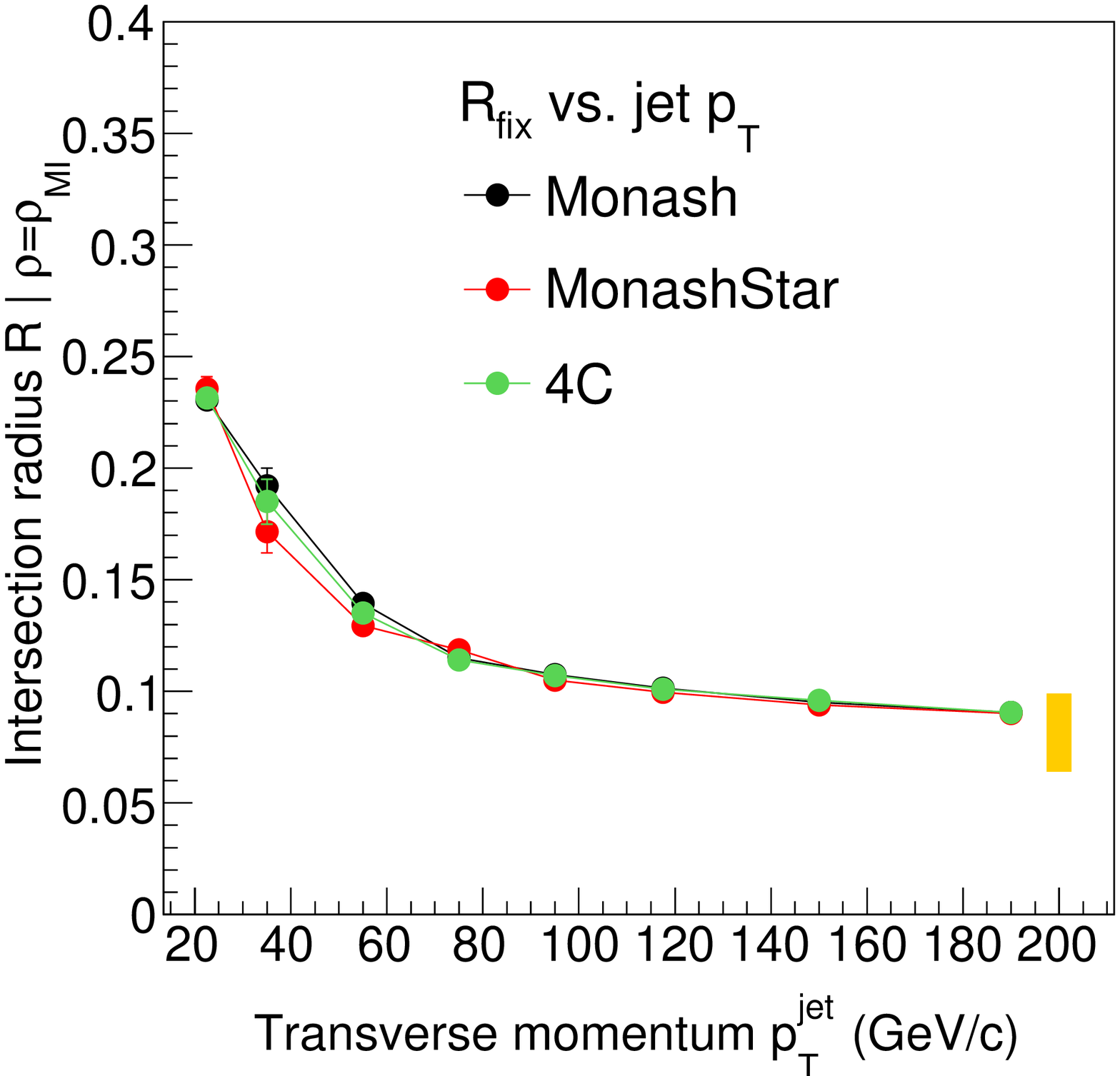}
\caption{\label{fig:rfix1}({\bf{Left}}) Ratio $\rho/\rho_\mathrm{MI}$ of the differential jet structure over the multiplicity-integrated curve for many different multiplicity bins; ({\bf{Right}}) The $p_\mathrm{T}^{\mathrm{jet}}$ dependence of $R_{\mathrm{fix}}$ with $\delta r = 0.1$. The~systematic uncertainty from the choice of $\delta r$ is shown as the yellow~band.}
\end{figure}

We also investigated whether the choice of multiplicity bins affects the location of the intersection point by categorizing the events into several smaller event multiplicity bins. We found that all the $\rho/\rho_\mathrm{MI}$ curves corresponding to these different multiplicity classes intersect each other at approximately the same radius, as~can be seen on the left panel of Figure~\ref{fig:rfix1}. Therefore we named this specific $r$ value $R_{\mathrm{fix}}$. The~$R_{\mathrm{fix}}$ depends, however, on~the transverse momentum of the jet, as~shown in the right panel of Figure~\ref{fig:rfix1}. The~$R_\mathrm{fix}(p_\mathrm{T}^\mathrm{jet})$ curve goes in a fashion expected by a Lorentz boost and converges to a constant value at higher energies~\cite{VVB}.

\begin{figure}[H]
\centering
\includegraphics[width=0.48\linewidth]{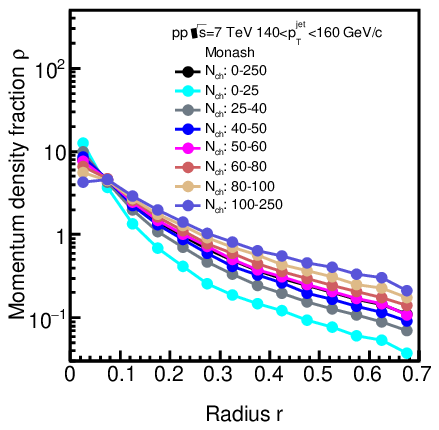}
\includegraphics[width=0.48\linewidth]{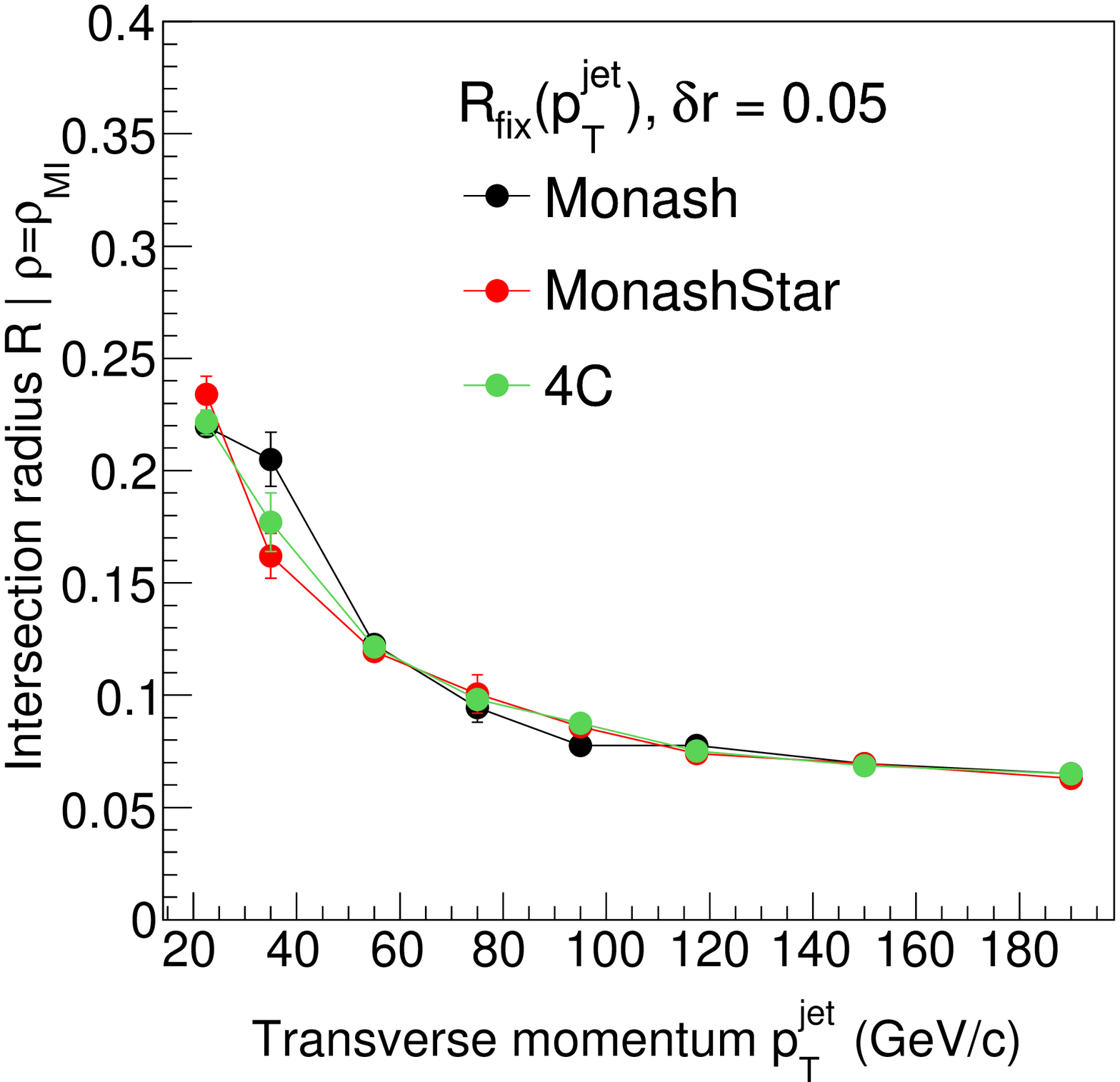}
\caption{\label{fig:rfix2}({\bf{Left}}) The differential jet shape for many different multiplicity bins with finer ($\delta r = 0.05$) binnings; ({\bf{Right}}) The $p_\mathrm{T}^{\mathrm{jet}}$ dependence of $R_{\mathrm{fix}}$ with finer ($\delta r = 0.05$) binnings.}
\end{figure}

Since we use a linear interpolation between the points of $\rho/\rho_\mathrm{MI}$, the~value of $R_{\mathrm{fix}}$ will have a slight dependence on the bin width in $r$ ($\delta r$), especially at higher $p_\mathrm{T}$ values where the value of $R_{\mathrm{fix}}$ is smaller. To~make sure that the results are robust enough, we repeated the analysis with narrower bins ($\delta r = 0.05$). In~the left panel of Figure~\ref{fig:rfix2}, we show $\rho$ with this finer $r$ binning, while in the right panel of the same figure we plotted the $p_\mathrm{T}^{\mathrm{jet}}$ dependence of $R_{\mathrm{fix}}$. We can see that the effects by the choice of bin width is rather small and does not change the conclusions, so we do not have to sacrifice the current statistics for finer~binnings.
\unskip
\unskip

\begin{figure}[H]
\centering
\includegraphics[width=0.48\linewidth]{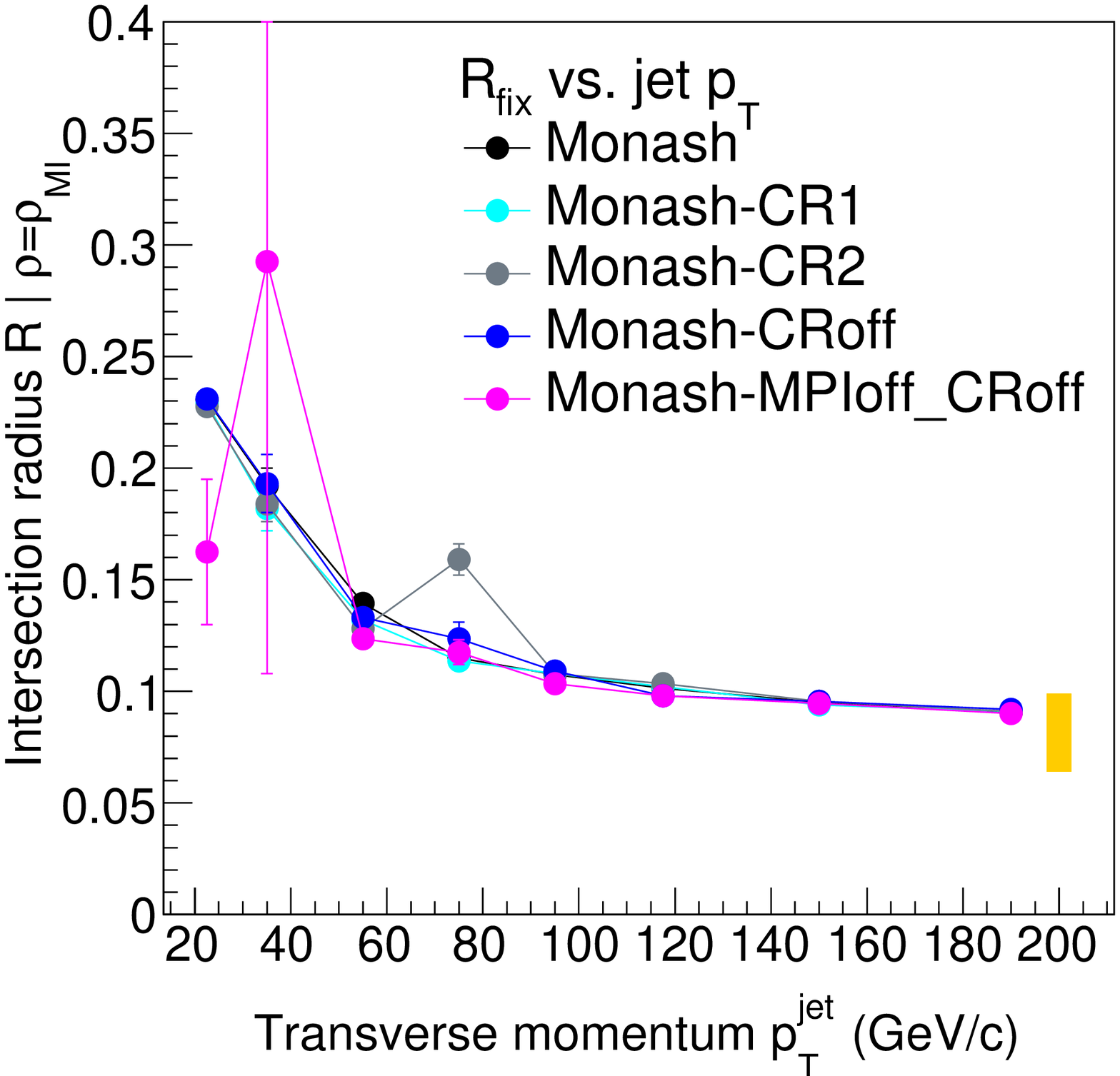}
\includegraphics[width=0.48\linewidth]{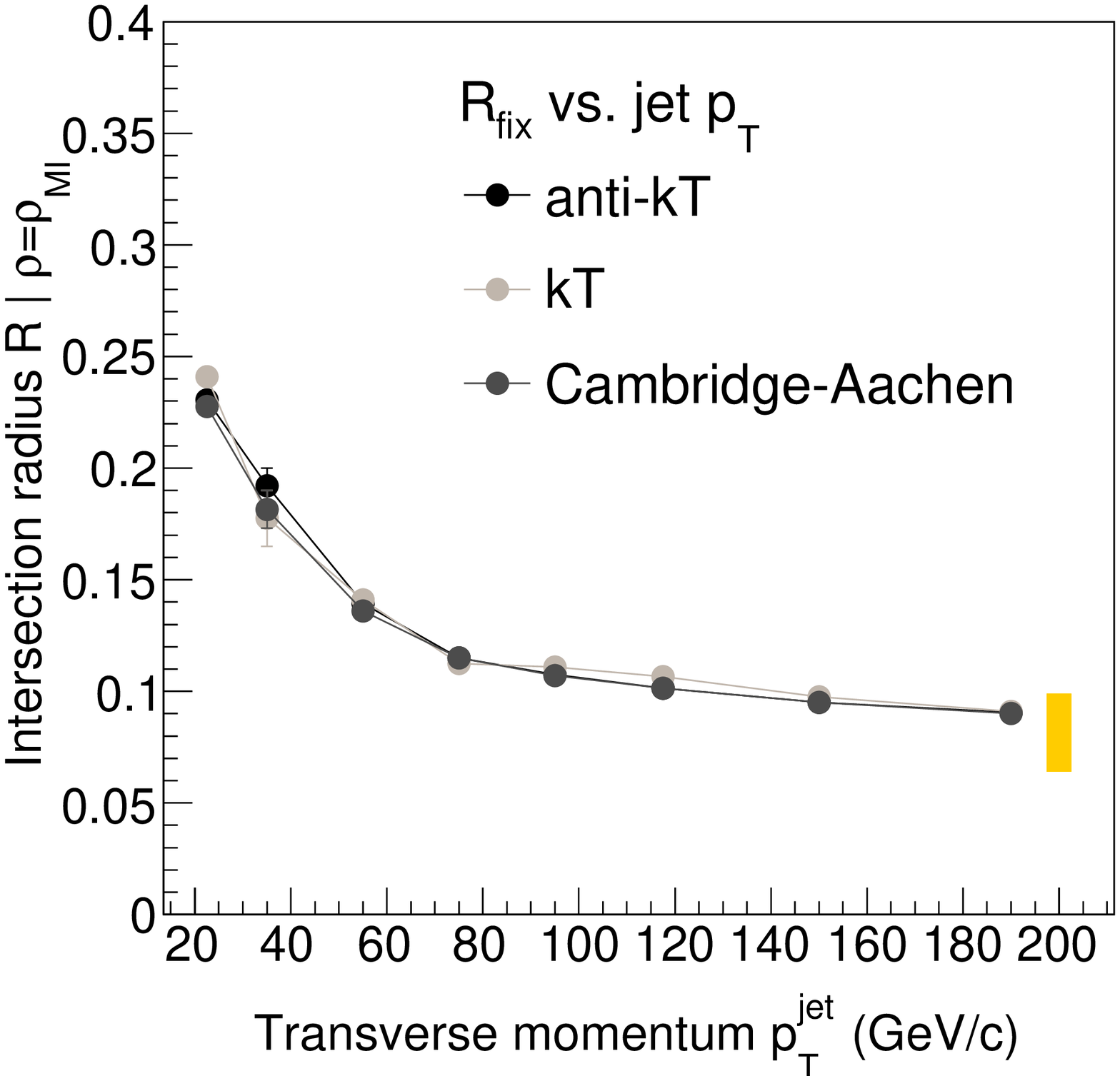}
\caption{\label{fig:rfix3} The evolution of $R_{\mathrm{fix}}$ with the $p_\mathrm{T}^{\mathrm{jet}}$ for different settings ({\bf{left}}); and~for different jet reconstruction algorithms ({\bf{right}}).}
\end{figure}

Using different settings in {\sc Pythia} changes the physics enough to expect different jet structures after jet reconstruction. All of the $k_\mathrm{T}$, Cambridge--Aachen and anti-$k_\mathrm{T}$ jet clustering algorithms that we investigated reconstructed the jet structures differently since they had different susceptibility to the underlying event. We therefore investigated the effects of varying the physics settings and used three different jet reconstruction algorithms (Figure \ref{fig:rfix3}). We conclude that the presence of $R_\mathrm{fix}$ was very robust for the different physics selections, as~its stability was neither an artifact of the particular choice of jet reconstruction algorithms, nor did it depend strongly on the underlying event. We note that $R_\mathrm{fix}$ is localized to lower $r$ values, therefore we do not expect significant influence from effects that are mostly visible at higher $r$ values, such as the modification of the UE or higher order corrections to the parton shower~\cite{Hoeche:NLL, Cal:NLL}.
We also computed the differential jet shapes using a different MC generator. We selected {\sc Hijing++} for this purpose because it implements a mechanism of creating the underlying event and QCD effects on the soft-hard boundary that is different from {\sc Pythia}. {\sc Hijing++} uses the {\sc Pythia} jet fragmentation, therefore we do not expect any difference during the later stages. Instead of MPI as implemented in {\sc Pythia}, {\sc Hijing++} uses minijet production. Differences at lower momenta may arise below the minijet cutoff. In~case of the $p_\mathrm{T}$ and multiplicity distributions, these effects do not exceed the variation caused by applying different tunes in {\sc Pythia}.

\begin{figure}[H]
\centering
\includegraphics[width=0.48\linewidth]{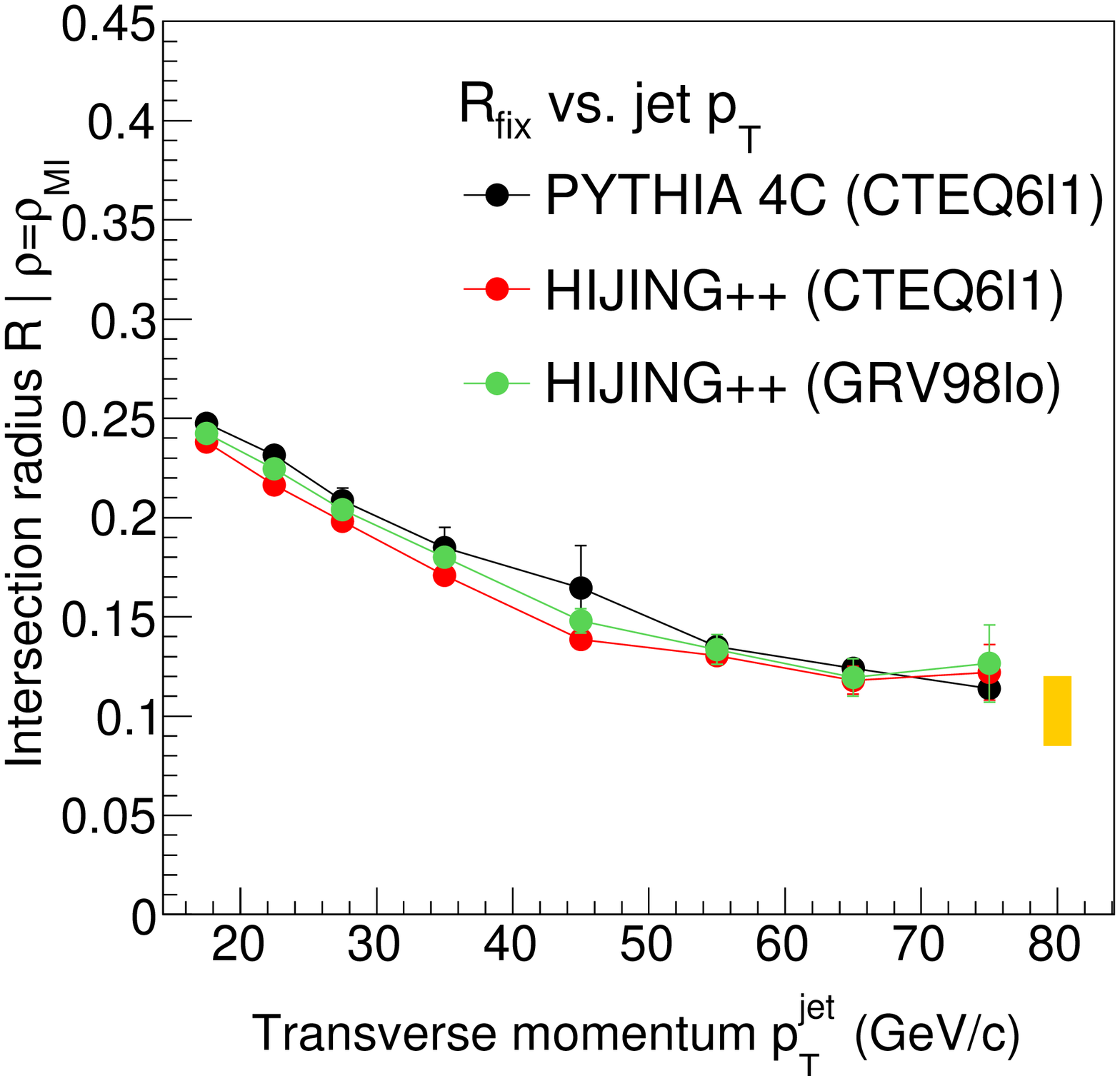}
\includegraphics[width=0.48\linewidth]{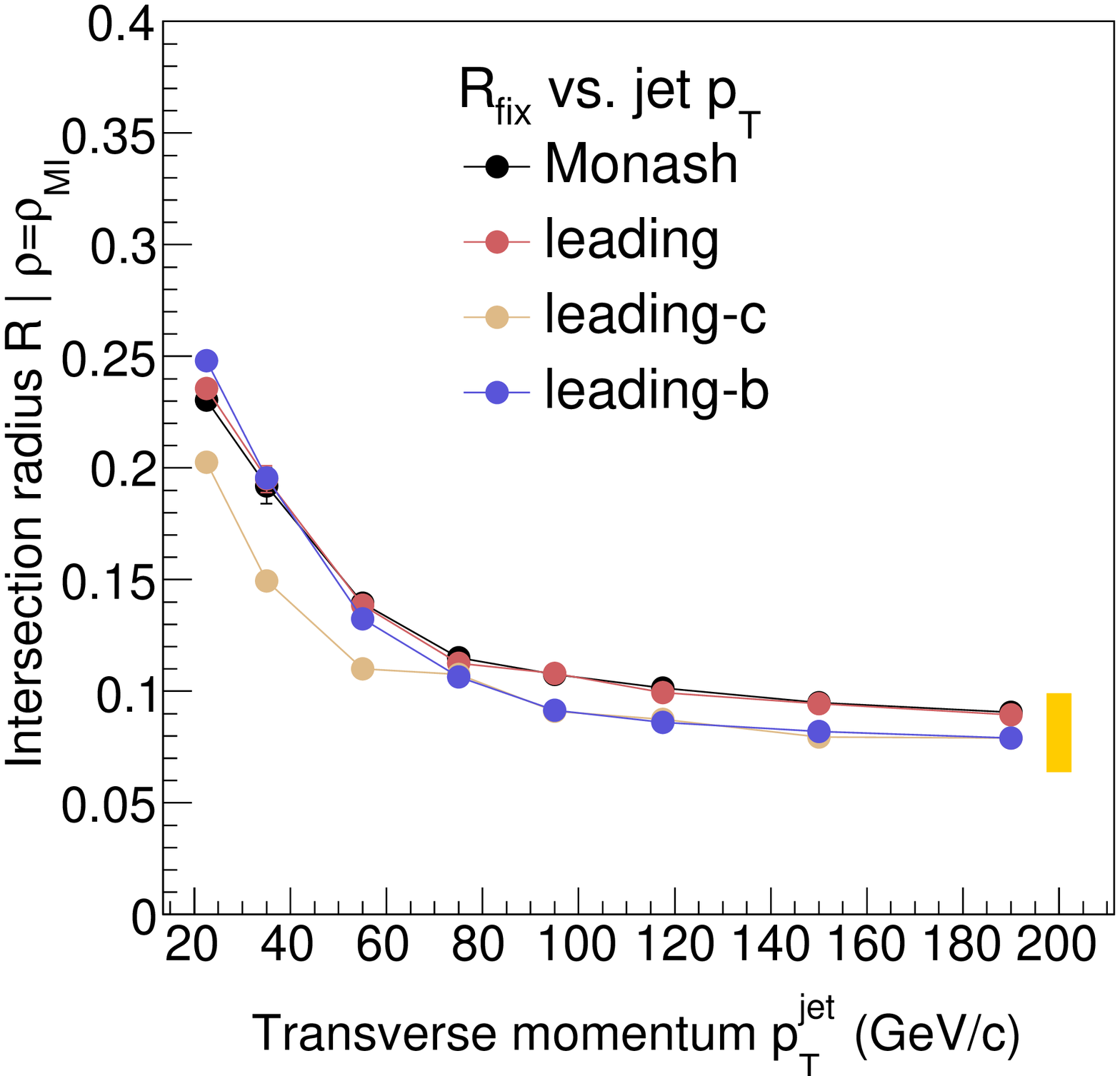}
\caption{\label{fig:rfix4} ({\bf{Left}}) The $p_\mathrm{T}^{\mathrm{jet}}$ dependence of $R_{\mathrm{fix}}$ for the {\sc Pythia} 4C tune is compared with {\sc Hijing++}; ({\bf{Right}}) Comparison of different jet selections for the {\sc Pythia} Monash~tune.}
\end{figure}

In the case of {\sc Hijing++}, we used two different PDF
sets. The~results show quantitatively the same $R_\mathrm{fix}$ dependence on $p_\mathrm{T}^\mathrm{jet}$ within systematic errors, as~can be seen in the left panel of Figure~\ref{fig:rfix4}. Jets originating from different flavors undergo different fragmentation due to both the color-charge effect and the dead-cone effect~\cite{deadcone}. We compared flavor-inclusive jets to heavy-flavor (beauty and charm) jets in the right panel of Figure~\ref{fig:rfix4}. We ensured that heavy flavor comes from the initial stages by only enabling leading order processes in PYTHIA. We compared these to leading and subleading flavor-inclusive~jets. 

It should be noted that the effect of non-leading jets is negligible on the $R_\mathrm{fix}$. Although~the overall tendency of heavy-flavor $R_\mathrm{fix}$ is similar to that observed for light flavor, there is also a clear quantitative difference between heavy and light flavors, which points to a different jet structure. The~leading b
jets differ for higher $p_\mathrm{T}^{\mathrm{jet}}$ and the leading c jets for lower $p_\mathrm{T}^{\mathrm{jet}}$. This suggests that the interplay between the mass and color-charge effects is non-trivial and needs further investigation. One possibility for that would be a parallel study of the UE and the fragmentation region  corresponding to a heavy-flavor trigger, in~a similar manner to~\cite{Ortiz:2018vgc}.
From all the above, we can assume that $R_\mathrm{fix}$ is a property of the jets that is associated with the final~state.

We also repeated the same analysis by selecting only the leading and sub-leading jets. We observed no significant difference compared to the case where all the jets were used (right panel of Figure~\ref{fig:rfix4}).

\unskip

\section{Conclusions}

We conducted a systematic study on jet structures for pp collisions using MC generators {\sc Pythia} and {\sc Hijing++}. We investigated the effects of CR and MPI on jet shapes and showed that not considering the effects of CR and MPI causes significant jet shape modification. We introduced a characteristic jet size ($R_\mathrm{fix}$) that depends only on the $p_\mathrm{T}$ of the jets but is independent of the jet reconstruction algorithms, MC generators, parton density functions, and~even  the choice of simulation parameters such as CR and MPI~\cite{VVB, VVB2}. We have also shown that the choice of $\delta r$ does not change our conclusions about $R_\mathrm{fix}$. These observations suggest that $R_\mathrm{fix}$ is an inherent property of the jets and is characteristic to the 
space-time evolution
of the parton shower at a given~momentum.

However, $R_\mathrm{fix}$ does depend on the flavor of the jet. Flavor-dependent jet structure studies may be a way to access mass versus color charge effects that is complimentary to hadron- or jet-production cross-section measurements. We believe that our findings motivate further phenomenology studies as well as cross-checks with real data to gain a deeper understanding on flavor-dependent jet fragmentation. Another direction for future research could be to investigate the effects of  MPI on jets without the underlying event. This could be done either by choosing an observable that depends very weakly on the underlying event~\cite{4jet}, or~by the parallel understanding of the underlying event and the fragmentation region~\cite{Ortiz:2018vgc}.

\vspace{6pt} 



\authorcontributions{Software and formal analysis, Z.V., R.V.; investigation, R.V., Z.V., G.G.B.; writing—original draft preparation, Z.V.; writing—review and editing, R.V.; supervison, R.V.; funding acquisition, G.G.B.}

\funding{This work has been supported by the NKFIH/OTKA K 120660 grant, the~János Bolyai scholarship of the Hungarian Academy of Sciences (R.V.), and~the MOST-MTA Chinese-Hungarian Research Collaboration. The~work has been performed in the framework of COST Action CA15213 THOR.}

\acknowledgments{The authors are thankful for the many useful conversations they had with Gábor Bíró, Jana Bielčíková, Miklós Kovács, Filip K\v{r}\'\i{}\v{z}ek, Yaxian Mao, Antonio~Ortiz and Guy Paić.}

\conflictsofinterest{The authors declare no conflict of~interest.} 

\reftitle{References}

\end{document}